\documentclass[prl,twocolumn,floatfix,superscriptaddress]{revtex4-1}
\usepackage{dcolumn,amsmath}
\usepackage{graphicx}
\usepackage{bm}
\usepackage{hyperref}

\newcommand{\eEDM}{{\em e}EDM}

\newcommand{\Eeff}{\ensuremath{E_{\rm eff}} \ }
\begin{document}

 \title{Electric-field-dependent $g$ factor for the ground state of lead monofluoride, PbF}
\author{V.V. Baturo}
\affiliation{Saint Petersburg State University, St. Petersburg, 199034,
Russia}
\affiliation{Federal State Budgetary Institution ``Petersburg Nuclear Physics  Institute'', Gatchina, Leningrad District 188300, Russia}
\author{P. M. Rupasinghe}\affiliation{Department of Physics, State University of New York at Oswego, Oswego, New York 13126, USA}
\author{T. J. Sears}\affiliation{Department of Chemistry, Stony Brook University, Stony Brook, New York 11794-3400, USA}
\author{R. J.\ Mawhorter}
\affiliation
{Department of Physics and Astronomy, Pomona College, Claremont, California 91711, USA}
\author{J.-U. Grabow}\affiliation{Gottfried-Wilhelm-Leibniz-Universit\"{a}t, Institut f\"{u}r Physikalische Chemie and Elektrochemie, Lehrgebiet A, D-30167 Hannover, Germany}
\author{A.N.\ Petrov}\email{petrov\_an@pnpi.nrcki.ru}
\homepage{http://www.qchem.pnpi.spb.ru}
\affiliation{Saint Petersburg State University, St. Petersburg, 199034,
Russia}
\affiliation{Federal State Budgetary Institution ``Petersburg Nuclear Physics  Institute'', Gatchina, Leningrad District 188300, Russia}

\begin{abstract}
The electric-field-dependent $g$ factor and the  electron electric dipole moment (eEDM)-induced Stark splittings for the lowest rotational levels of $^{207,208}$PbF are calculated. Observed and calculated Zeeman shifts for $^{207}$PbF are found to be in very good agreement. It is shown that the $^{207}$PbF hyperfine sublevels provide a promising system for the eEDM search and related experiments.
\end{abstract}

\maketitle



 The spectroscopic and theoretical work on the PbF molecule over more than three decades including\cite{Kozlov:87, Ziebarth:98, Dmitriev:92, Shafer-Ray:06, Shafer-Ray:08E, Baklanov:10, Mawhorter:11, Alphei:11, Yang:13, Petrov:13, Skripnikov:15d} has been reported.   Based on data at optical resolution \cite{Ziebarth:98}, Shafer-Ray et al.\cite{Shafer-Ray:06} predicted that the electric-field-dependent $g$ factor of the ground state of $^{208}$PbF could cross zero at an electric field of 68 kV/cm.  This led to the conclusion that PbF might provide a uniquely sensitive probe of the electric dipole moment of the electron (eEDM), $d_e$.\\

Working to verify this, subsequent spectroscopy at a higher resolution by McRaven et al.\cite{Shafer-Ray:08E} and a theoretical analysis in Ref.\cite{Baklanov:10} revealed a misassignment of the parity of the lowest rotational levels and confusion concerning the sign of the large $^{207}$Pb Frosch-Foley d (= -$A_{\perp}$) hyperfine parameter in the optical work on $^{207}$PbF\cite{Ziebarth:98, Mawhorter:11}.  The reanalysis performed by Yang et al.\cite{Yang:13} with corrected spectroscopic constants shown the $g$ factor of the ground-state $^{208}$PbF unfortunately does not vanish. Nevertheless the very small $g$ factor in the $^2\Pi_{1/2}$ ground state of PbF reduces the sensitivity to stray magnetic fields by about a factor of 20 with regard to comparable $^2\Sigma$ molecules. This is a significant advantage in parity nonconservation studies. \\

Analytical expressions for the electric-field-dependent $g$ factor were obtained\cite{Shafer-Ray:06, Yang:13} for $^{208}$PbF under the assumption that the mixing of different rotational levels by an electric field is not important.  In this article  we take the mixing into account by the numerical inclusion of a large number of rotational states and consider both odd and even mass Pb isotopologues of PbF.  \\

$^{208}$Pb is the most abundant lead $I=0$ isotope with 52\% natural abundance, while $^{207}$Pb has a nuclear spin $I=1/2$ and a natural abundance of 22\%.  The existence of the lead nuclear spin in $^{207}$PbF has a surprisingly strong effect on the Zeeman splittings in low-lying fine and hyperfine split levels that has major implications  for experimental \eEDM\, searches.  It was shown by Alphei et al.\cite{Alphei:11} that a coincidental near-degeneracy of levels of opposite parity in the ground rotational state $J=1/2$ for $^{207}$PbF\cite{Shafer-Ray:08E, Alphei:10} takes place, caused by the near cancellation of energy shifts due to $\Omega$-type doubling and the $^{207}$Pb$^{19}$F magnetic hyperfine interactions. Thus $^{207}$PbF has also been proposed as a promising candidate for both temporal variation of the fundamental constants \cite{Flambaum:2013} and anapole moment \cite{Alphei:11, Borschevsky:13} experiments .  \\

 This general utility of $^{207}$PbF for a variety of parity non-conservation experiments offers an alternative path to spectroscopically probing states of different parity and can be further enhanced by working with excited vibrational levels in the ground electronic $X_1$ state \cite{MAWHORTER:18}.  The levels of opposite parity ($^{207}$PbF levels 3 and 4 in Fig. 1 of Ref. \cite{Mawhorter:11}) are only 266 MHz apart for v=0 and drop about 33 MHz for each step up the vibrational ladder, potentially crossing with a gap of only $\sim \pm$20 MHz around $v=7$ and $v=8$. \\

The knowledge of $g$ factors helps to control and suppress important systematic effects due to stray magnetic fields \cite{Petrov:14, Petrov:17b, ACME:18}. However, neither theoretical nor experimental data for $g$ factors of $^{207}$Pb$^{19}$F for the field-free case or in an external electric field have been reported to date. The main purpose of the article is to fill this gap.

\section{Experimental Details}
As described in detail in our earlier study\cite{Mawhorter:11}, the rotational Zeeman spectra were taken at the Gottfried-Wilhelm-Leibniz-Universität Hannover using a Fourier-transform microwave (FTMW) spectrometer that exploits a coaxial arrangement of the supersonic jet and resonator axes (COBRA)\cite{Grabow;96}. The resulting sensitivity coupled with the laser ablation\cite{Giuliano;08} of elemental Pb in a neon carrier gas augmented with a few percent of SF$_{6}$ enabled the observation of strong and robust signals, which were essential to measuring the Zeeman effect data for both $^{208}$PbF and $^{207}$PbF. As already mentioned, due to a cancellation of spin and orbital contributions inherent in the 


\begin{figure*}
	\centering
	\includegraphics[width=0.85\linewidth]{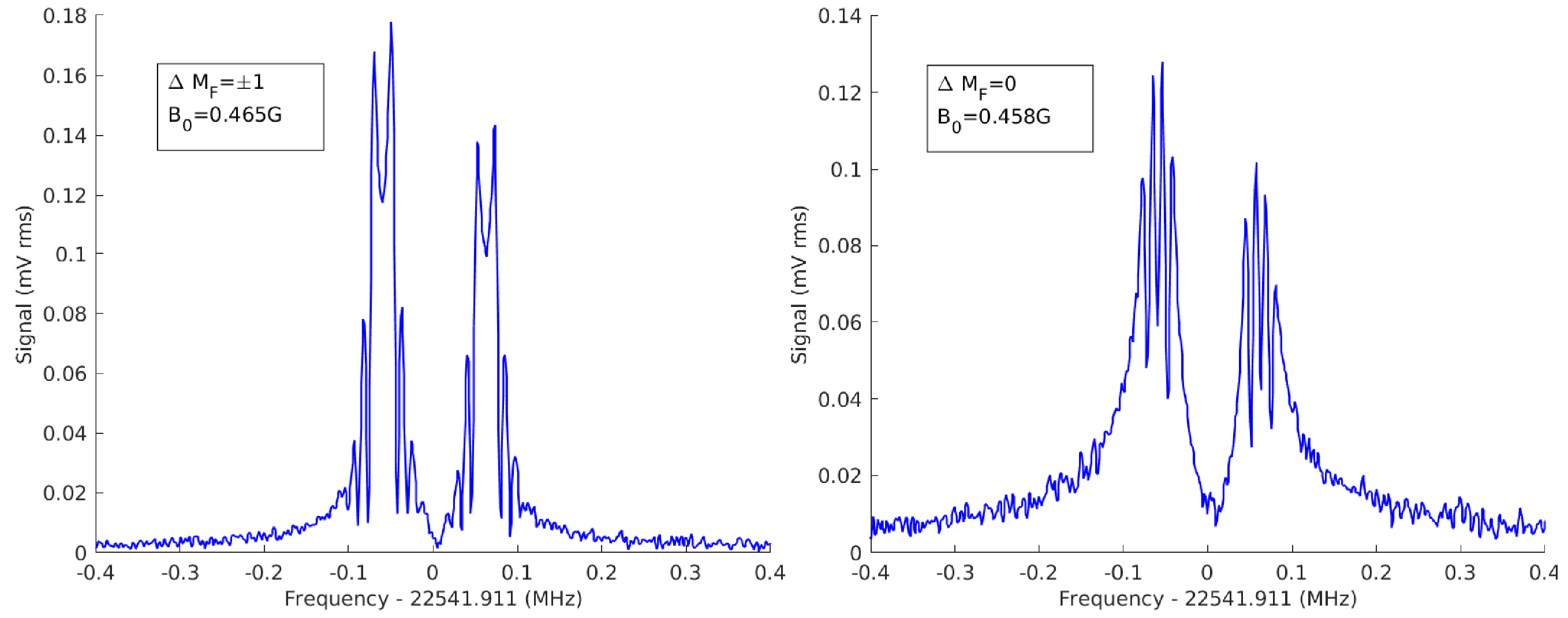}
	\caption{Rotational Zeeman spectra for the $^{207}$PbF $F_L=3/2 \rightarrow F_U=5/2$ transition at 22 541.912 MHz, showing the comparable (left) eight $\Delta M_F = \pm 1$ (B$_0$ = 0.465 G) and (right) four $\Delta M_F = 0$ (B$_0$ = 0.458 G) splittings, respectively. Note the doubled transitions due to the COBRA Doppler effect.}
	\label{fig:twopanelfig2}
\end{figure*}

\begin{figure}
	\centering
	\includegraphics[width=0.85\linewidth]{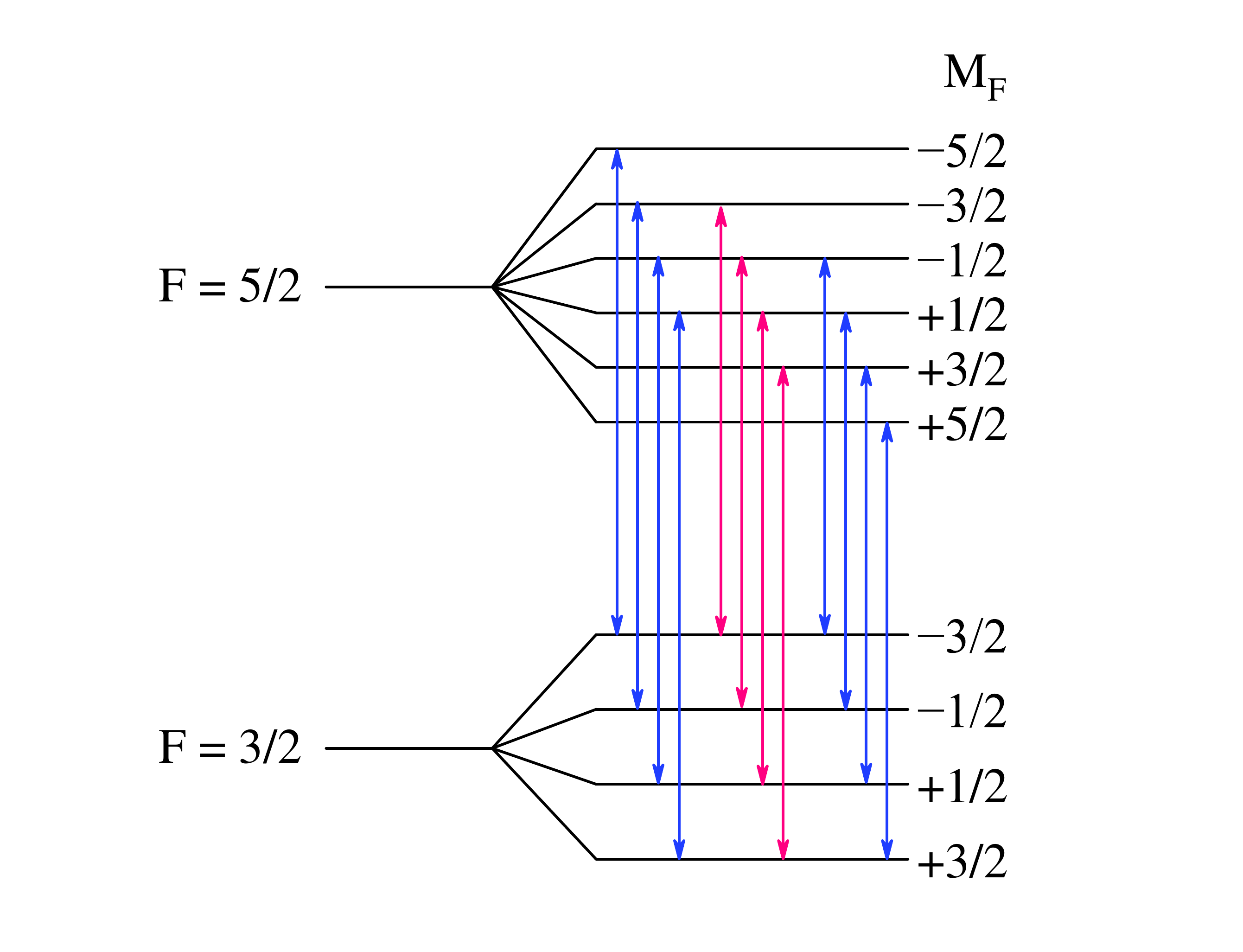}
	\caption{ A depiction of rotational Zeeman spectra shown in Fig. \ref{fig:twopanelfig2}.
	The $\Delta M_F = \pm 1$ are marked by blue arrows The $\Delta M_F = 0$ components are marked by red arrows.}
	\label{level_diagram}
\end{figure}

\noindent$^2\Pi$ ground state of PbF, the observed Zeeman splittings are small.  Even so, the excellent signal to noise with long emission decay times allows frequency measurements for unblended lines at an accuracy of ~0.5 kHz and the resolution of transitions separated by more than 6 kHz.  Figure \ref{fig:twopanelfig2} shows representative Zeeman spectra for the 22541.912 MHz transition in $^{207}$PbF. Note that the resonance signals are doubled due to the velocity structure in the experimental design. 
Figure \ref{level_diagram} shows the transitions contributing to the spectra. 
\\ 

\begingroup
\begin{table*}
	\caption{Observed  ($\Delta \nu/B)_{expt}$ (MHz/G) and calculated ($\Delta \nu/B)_{th}$ (MHz/G) Zeeman shifts of the $J=1/2 \rightarrow J=3/2$ transitions for $^{207}$PbF.
$\Delta \nu$ is the difference between transition frequencies when a magnetic field of value $B$ is applied and a field-free case.
		The number in parentheses gives a two standard deviation error in the final digits of precision. The subscripts $L$ and $U$ in $F_L, F_U, M F_L, M F_U$ refer to the upper and lower  energy level  of the transition, respectively. $F$ means the total (electronic  plus rotational plus nuclear spins) angular momentum, and $MF$ its projection to the laboratory axis. }
	
\begin{ruledtabular}
	\begin{tabular}{dccddddd}
		\multicolumn{1}{c}{Unsplit line (MHz)} &\multicolumn{1}{c}{$F_L$} &\multicolumn{1}{c}{$F_U$} &\multicolumn{1}{c}{$M F_L$}  &\multicolumn{1}{c}{$M F_U$} &\multicolumn{1}{c}{$(\Delta \nu/B)_{exp}$}  & \multicolumn{1}{c}{$(\Delta \nu/B)_{th}$} & \multicolumn{1}{c}{$(\Delta \nu/B)_{exp}-(\Delta \nu/B)_{th}$ }\\
		\hline
		18333.501 & 3/2 & 5/2 & 3/2	& 1/2
		& -0.101(12)& -0.1121(60) & 0.0111 \\
		& & & 1/2 & -1/2 &-0.0755(30)&-0.0803(28)&  0.0048 \\
		& & & -1/2 &-3/2 &-0.04984(45) &-0.0486(51) & -0.0012\\
		& & & 3/2 &	3/2 &-0.0453(20) &-0.0475(83) & 0.0022\\
		& & & -3/2&-5/2 &-0.01650(48) &-0.017(11) & 0.001\\
		& & & 1/2&1/2&-0.0139(20) &-0.0157(27)&  0.0018\\
		& & & -1/2&-1/2&0.0159(20) &0.0160(28) &  -0.0001\\
		& & & 3/2&5/2&0.01647(50) &0.017(11) &  -0.001\\
		& & &-3/2&-3/2&0.0474(20)	&0.0475(83)&  -0.0001\\
		& & & 1/2&3/2&0.05035(37)	&0.0489(50)&  0.0015\\
		& & & -1/2&1/2&0.0763(47)	&0.0805(28)&  -0.0042\\
		22541.912  &3/2 &5/2 &-3/2&	-1/2&	-0.0957(35) &	-0.0910(59)&	-0.0047\\
		& & & -1/2&	1/2	&-0.0689(21)&	-0.0693(27)	&0.0004\\
		& & & 1/2&	3/2&	-0.04727(76)&	-0.0476(50)&0.0003	\\
		& & & -3/2&	-3/2&	-0.03326(83)&	-0.0326(81)&	-0.0007	\\
		& & & 3/2&	5/2&	-0.02523(52)&	-0.026(10)&	0.001	\\
		& & & -1/2&	-1/2&	-0.01093(30)&	-0.0109(27)&	0.0000	\\
		& & & 1/2&	1/2&	0.01062(48)&	0.0109(28)&	-0.0003	\\
		& & & -3/2&	-5/2&	0.02504(57)&	0.026(10)	&-0.001	\\
		& & & 3/2&	3/2&	0.03316(76)&	0.0326(82)&	0.0006\\
		& & &-1/2	&-3/2&	0.0466(10)&	0.0475(49)&	-0.0009	\\
		& & & 1/2&	-1/2&	0.0675(26)&	0.0692(27)& -0.0017	\\
		& & & 3/2&	1/2&	0.0983(58)&	0.0910(60)& 0.0073	\\
		22658.902 &1/2 & 3/2 &-1/2&	-3/2&	-0.1448(10)&	-0.1465(50)&	0.0017	\\
		& & &-1/2&	-1/2&	-0.05157(74)\footnotemark[1]&	-0.0500(17)	&-0.0016	\\
		& & &1/2&	-1/2&	-0.0507(25)\footnotemark[1] &	-0.0470(17)&	-0.0037	\\
		& & &-1/2&	1/2&	0.05073(99)&	0.0467(16)&0.0040	\\
		& & & 1/2 &1/2&	0.0516(19)&0.0497(16)	&0.0019		\\
		& & & 1/2 &3/2&0.14387(52)&	0.1465(49)&	-0.0026	\\
		22691.749 &1/2 &1/2 & 1/2&	-1/2&	-0.09555(46)& -0.1000(33)&	0.0045	\\
		& & &-1/2	&1/2	&0.09819(81)	&0.1003(32)	&-0.0021	\\
	\end{tabular}
\end{ruledtabular}
\footnotetext[1]{Typographic error in \cite{Baum:10}, corrected here}
\label{Zeman207}
\end{table*}
\endgroup

\begingroup
\begin{figure}
\centering
\includegraphics[width=0.95\linewidth]{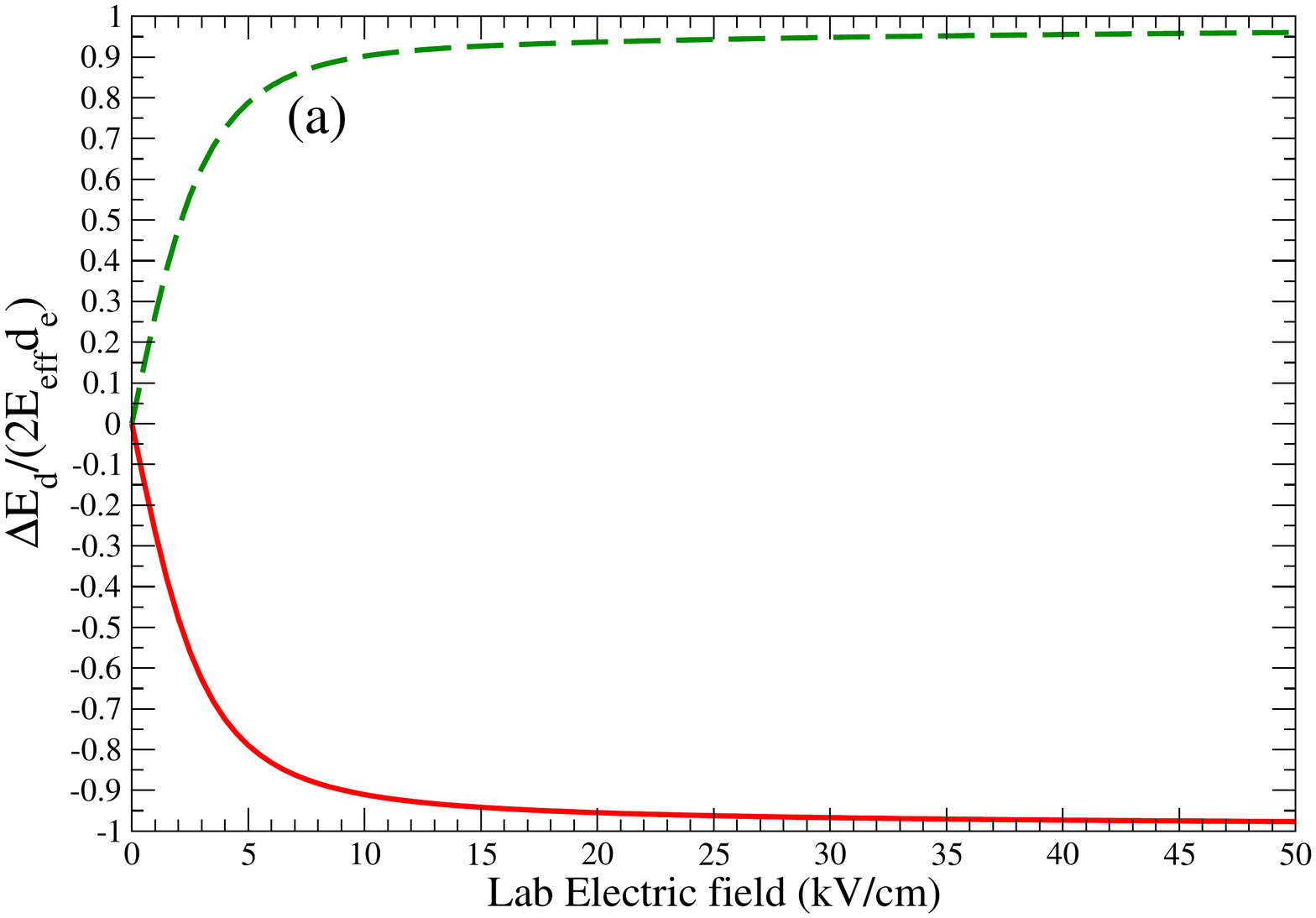}
\includegraphics[width=0.95\linewidth]{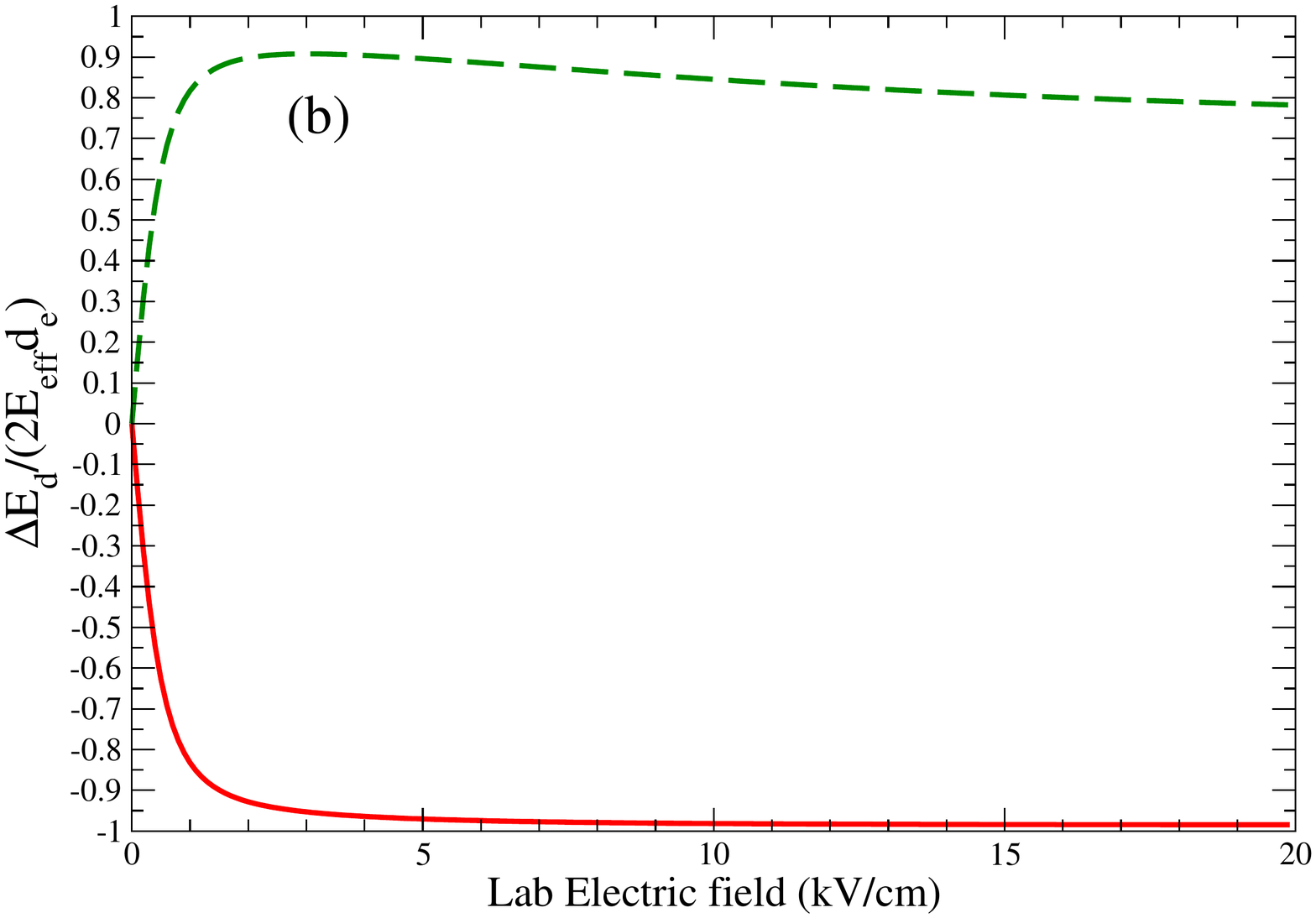}
\includegraphics[width=0.95\linewidth]{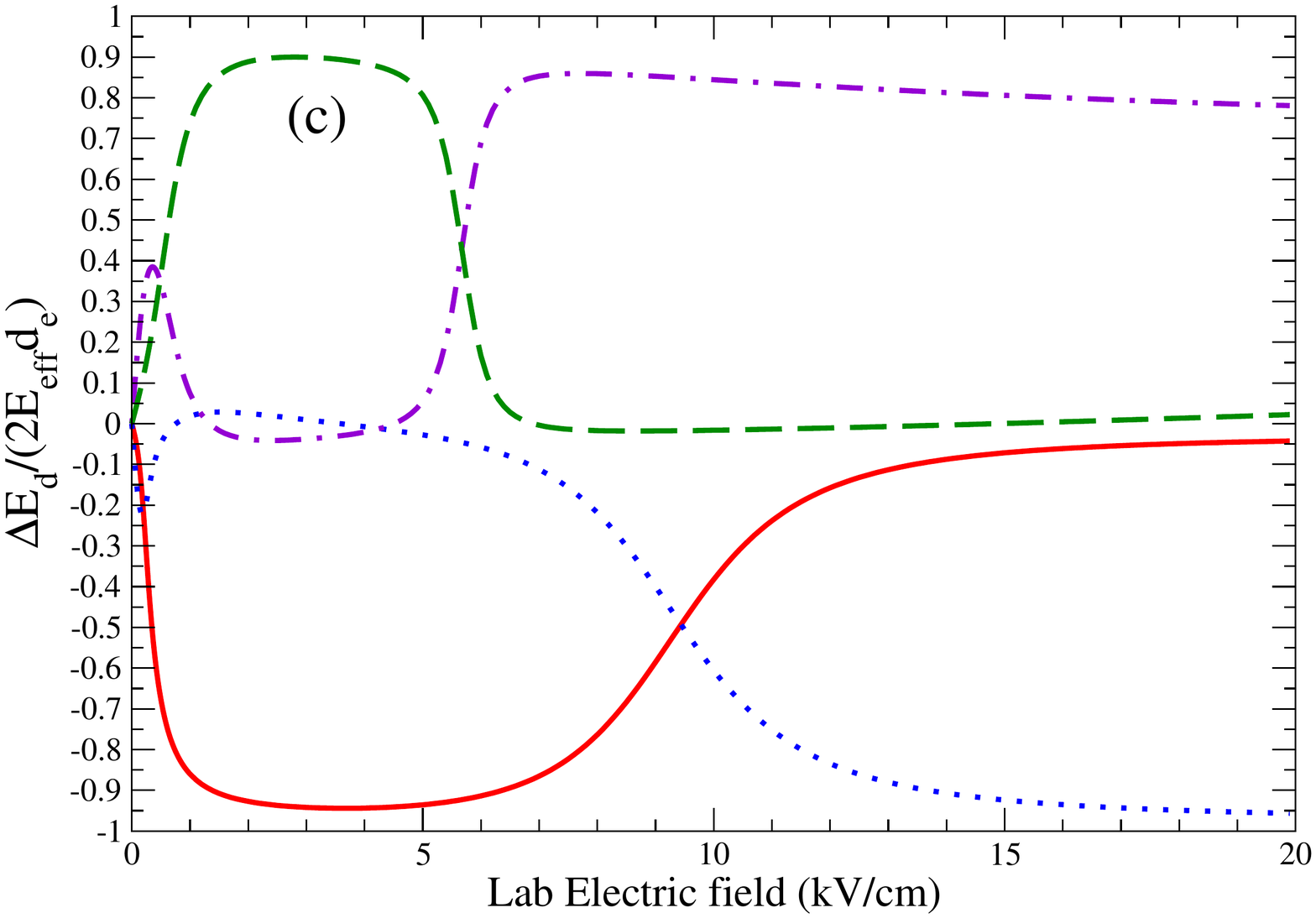}
\caption{\label{EDMsplitting} 
The eEDM-induced Stark splitting ($\Delta E$) between $\pm M_F$ pairs of
hyperfine states.
(a) $^{208}$PbF. The solid (red) line corresponds to the $|M_F|$=1 lower-lying
$\Omega$-doublet states whereas the dashed (green) line corresponds to the higher-lying $|M_F|$=1 states.
(b) $^{207}$PbF. The solid (red) line corresponds to the $|M_F|$=3/2 lower-lying
$\Omega$-doublet states whereas the dashed (green) line corresponds to the higher-lying $|M_F|$=3/2 states.
(c) $^{207}$PbF. The solid (red) line corresponds to the  lower-lying
$F$=3/2, $|M_F|$=1/2 $\Omega$-doublet states, the dashed (green) line corresponds to the higher-lying $F$=3/2, $|M_F|$=1/2 states, the dotted (blue) line corresponds to the lower-lying $F$=1/2, $|M_F|$=1/2 states,
 whereas the dashed-dotted (violet) line corresponds to the higher-lying $F$=1/2, $|M_F|$=1/2 states}
\end{figure}
\endgroup

Given the high resolution of the jet spectra, the magnetic field calibration becomes the primary factor determining the uncertainty of the molecule-fixed $g$ factor, $G_{\perp}$.  The currents in the three pairs of Helmholtz coils surrounding the chamber were independently varied to null out the magnetic field.  This was done by adjusting the Helmholtz coil currents for all three pairs until all Zeeman splittings are minimized. Having full three-axis control enabled the application of magnetic fields either perpendicular along two different axes or parallel to the radiation polarization. Having individual axis control, we can verify that the magnetic field in the sample region was indeed determined from the change in current in each coil. This was done by making independent experimental determinations of $G_{\perp}$, in both parallel and perpendicular configurations.  They agreed to within about 2.5\%, indicating that the uncertainty in our magnetic field calibration is approximately equal to the statistical error of our measurement. \\

Note that the initial experimental level assignments in Ref. \cite{Baum:10} have been reversed, resulting in the completely consistent set of experimental and theoretical $g$ factors presented here.  These are a factor of ~1.45 smaller than used in Refs. \cite{Mawhorter:11, Alphei:11}, and result in the prediction for the avoided level crossing discussed in Ref. \cite{Alphei:11} to occur at a magnetic field of approximately 1190 $\pm$ 80 G.

\section{Results and discussion}
The eigenvalues and eigenfunctions of the lead monofluoride molecule were obtained by numerical diagonalization of the Hamiltonian  over the basis set of the electronic-rotational and nuclear spin wave functions. Details of the method and parameters of the Hamiltonian can be found in Refs. \cite{Petrov:11, Petrov:13, Skripnikov:15d}. Parameters used include the body-fixed $g$ factors  $G_{\parallel}=0.081(5)$, $G_{\perp}=-0.27(1)$ \cite{Skripnikov:15d}, nuclear $g$ factors $g_{^{19}F}=5.25772 \mu_N$, $g_{^{207}Pb}=1.18204 \mu_N$ \cite{Fella:20}, and the body-fixed molecular dipole moment $D=1.38$ a.u. \cite{Mawhorter:11}.  Further details are provided in Refs. \cite{Mawhorter:11, rupasinghe;11}.

 The energy levels of interest for potential eEDM experiments on $^{208}$PbF are the $F^{p}=1^{-}$ and $F^{p}=1^{+}$ states which are the first and fourth energy levels in zero field. $p= \pm 1$ means the parity of a state.  For $^{207}$PbF, the levels of interest are the closely spaced $\Omega$-doublet states $F^{p}=3/2^{-}$, $F^{p}=1/2^{-}$, $F^{p}=1/2^{+}$, and $F^{p}=3/2^{+}$, which are the second, third, fourth and fifth energy levels. The relevant energy levels can be seen in Fig. 1 of Ref. \cite{Mawhorter:11}.
In an eEDM search experiment opposite parity levels are mixed in an electric field to polarize the molecule.  As the molecule becomes fully polarized the splitting $\Delta E_d$ between $\pm M_F$ levels due to an eEDM-related Stark shift reaches the maximum value $2d_eE_{\rm eff}$, where $\Eeff = 40\ {\rm GV/cm}$ \cite{Skripnikov:14c}  is the effective internal electric field. 
Assuming an eEDM value $|d_e| = 1.1\times 10^{-29}$ from the current limit \cite{ACME:18}  (ACME II experiment), we have $2d_eE_{\rm eff} = 0.2 {\rm mHz}$.
For any real electric field the splitting
is less than $2d_eE_{\rm eff}$ by an absolute value. In Fig. \ref{EDMsplitting} the calculated eEDM-induced Stark splittings for $^{208, 207}$PbF are presented.\\

The calculated and observed\cite{Baum:10} Zeeman shifts  of the $J=1/2 \rightarrow J=3/2$ transitions for $^{207}$PbF are given in Table \ref{Zeman207} and graphically in Fig. \ref{fig:Shifts}. The deviations between calculated and observed Zeeman shifts ($E_{\rm Z}$) are consistent with the estimated experimental and theoretical 
uncertainties ($\delta E_{\rm Z}$).
Conservative theoretical uncertainties were calculated as
\begin{equation} 
 \delta E_{\rm Z} = \sqrt{\left( \frac{\partial E_{\rm Z}}{\partial  G_{\parallel}} \delta G_{\parallel} \right)^2 +
 \left( \frac{\partial E_{\rm Z}}{\partial  G_{\perp}} \delta G_{\perp} \right)^2},
\end{equation} 
where $\delta G_{\parallel} = 0.005$, $\delta G_{\perp} = 0.01$ \cite{Skripnikov:15d}.

In the article we define the $g$ factors such that the Zeeman shift is equal to
\begin{equation} 
   E_{\rm Z} = g \mu_B B M_F,
 \label{Zeem}
\end{equation} 
where $M_F$ is the projection of the total angular momentum ${\bf F}$ (including nuclear spin) on the direction of $\mathbf{B}$ and the electric field, $\mathbf{E}$.
In  Fig. \ref{gfactor}, the calculated electric-field-dependent $g$ factors are presented. From Fig. (\ref{gfactor}) (a) one can see that taking into account the mixing of different rotational levels by the electric field is important for an accurate evaluation of the $g$ factors.\\

As can be seen in Fig. (\ref{EDMsplitting}) the advantage of the $^{207}$PbF molecule is that it is polarized at a lower electric field and has smaller absolute $g$ factors than does $^{208}$PbF. This is important for the eEDM experiment as larger fields and $g$ factors lead to greater systematic uncertainties in experimental measurements.  For $E=5$ kV/cm the eEDM Stark shift reaches 80\% of the maximum value for $^{208}$PbF, whereas for $^{207}$PbF, $|M_F|$=3/2 the same efficiency is achieved at $E=1$ kV/cm,  and for $E=2$ kV/cm it is 90\%. The values for the $g$ factors vary from $0.04$ to $0.01$.  As a comparative example, for the YbF molecule with $g=2$ the efficiency is only about 55\% for $E=10$ kV/cm \cite{Kara:2012}. \\

As a posited eEDM splitting $\Delta E$ between $\pm M_F$ levels is measured, the eEDM value
$d_e = \frac {\Delta E}{2E_{\rm eff}} $ can be extracted. However, according to Eq. (\ref{Zeem}), an external magnetic field also leads to a splitting, i.e., the assumed energy difference between the $+|M_F|$ and $-|M_F|$ levels
$\Delta E_Z = 2 g \mu_B B |M_F|$. Therefore a stray magnetic field leads to systematic effects, and the smaller is the g-factor the smaller are the corresponding systematics.
The complex hyperfine structure of $^{207}$PbF prevents a regular dependence on an electric field for both the eEDM Stark shift and the $g$ factor, as shown in Figs. \ref{EDMsplitting} and  \ref{gfactor}. There are several field values for which the $g$ factors are zero or near zero.  However, they are strongly (but not exactly) correlated with zero values for the eEDM Stark shift.
For example, for the higher-lying $F$=1/2, $|M_F|$=1/2 states
[the dashed-dotted (violet) line] the
$g$ factor is equal to zero at $E=0.87$ kV/cm,
 whereas $\Delta E_d = 0.25 d_eE_{\rm eff}$ is small (compared to the maximum value $ 2 d_eE_{\rm eff}$), but nonzero.

An efficient way to suppress the systematics related to the stray magnetic field is possible
if we have two different levels which have $\Delta E^1_Z = \Delta E^2_Z$ and opposite eEDM-induced splittings $\Delta E^1_d = - \Delta E^2_d$. In this case extracting the eEDM using the formula
$d_e = \frac {\Delta E^1 - \Delta E^2}{4E_{\rm eff}} $ will double the eEDM signal and cancel out the contribution from the stray magnetic field. Here $\Delta E^i = \Delta E^i_d + \Delta E^i_Z$ is the total splitting.
It has been shown previously that the levels with the required structure are closely spaced $\Omega$-doublet levels, such as in ThO \cite{ACME:18, DeMille:2001, Petrov:14, Vutha:2010, Petrov:15, Petrov:17} 
or HfF$^{+}$ \cite{Cornell:2017, Petrov:18}.
Unfortunately, Figs. \ref{EDMsplitting} and \ref{gfactor} show that the PbF molecule does not have levels with the required structure.
However, certain combinations of the splittings do allow for the cancellation of the Zeeman contribution while
keeping a nonzero contribution from the eEDM.
For example, if state ``1'' is the lower-lying
$F$=3/2, $|M_F|$=1/2 [solid (red) line]  and  state ``2'' is
the higher-lying $F$=3/2, $|M_F|$=1/2 [dashed (green) line],
then for $E=2$ kV/cm the combination of Zeeman splittings
$\Delta E^1_Z + 2.34 \Delta E^2_Z = 0$ cancels, while the 
contribution from the eEDM to the same combination
$\Delta E^1_d + 2.34 \Delta E^2_d = 2.3 d_eE_{\rm eff}$
is nonzero.


\begingroup
\begin{figure}
    \centering
    \includegraphics[scale=0.3]{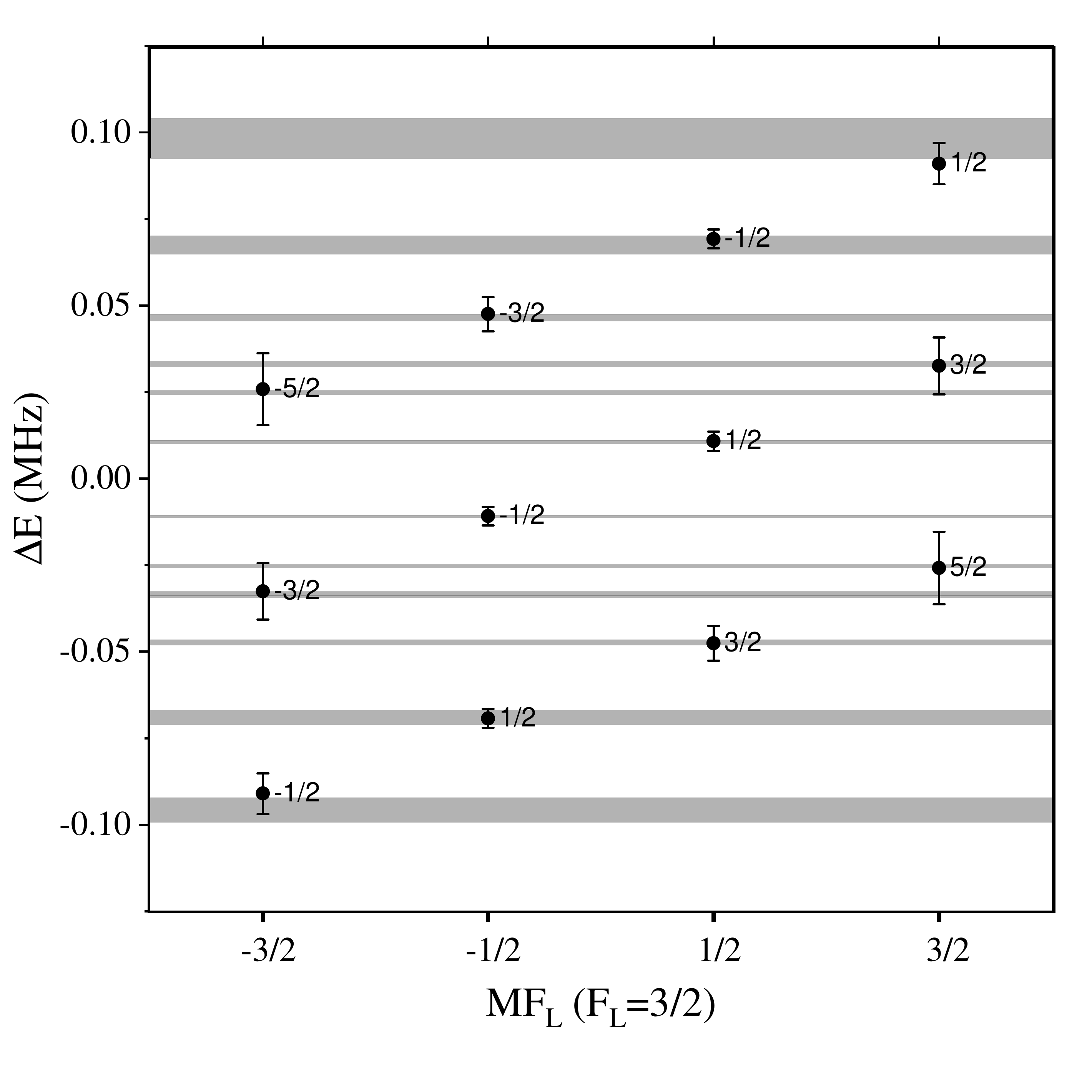}
    \caption{
Calculated (circles) and experimental (horizontal bands, bandwidths corresponding to two standard deviation uncertainty) shifts for the $F_L=3/2 \rightarrow F_U=5/2$ transition at $\nu = 22541.912$  MHz. $M F_L$ values are on the x-axis, and $M F_U$ values are marked in the figure. Note the excellent agreement and the eightfold/fourfold natures of the $\Delta M_F = \pm 1$ and $\Delta M_F = 0$ transitions clearly apparent in Fig. 1. }
    \label{fig:Shifts}
\end{figure}

\begin{figure}
\centering
\includegraphics[width=0.95\linewidth]{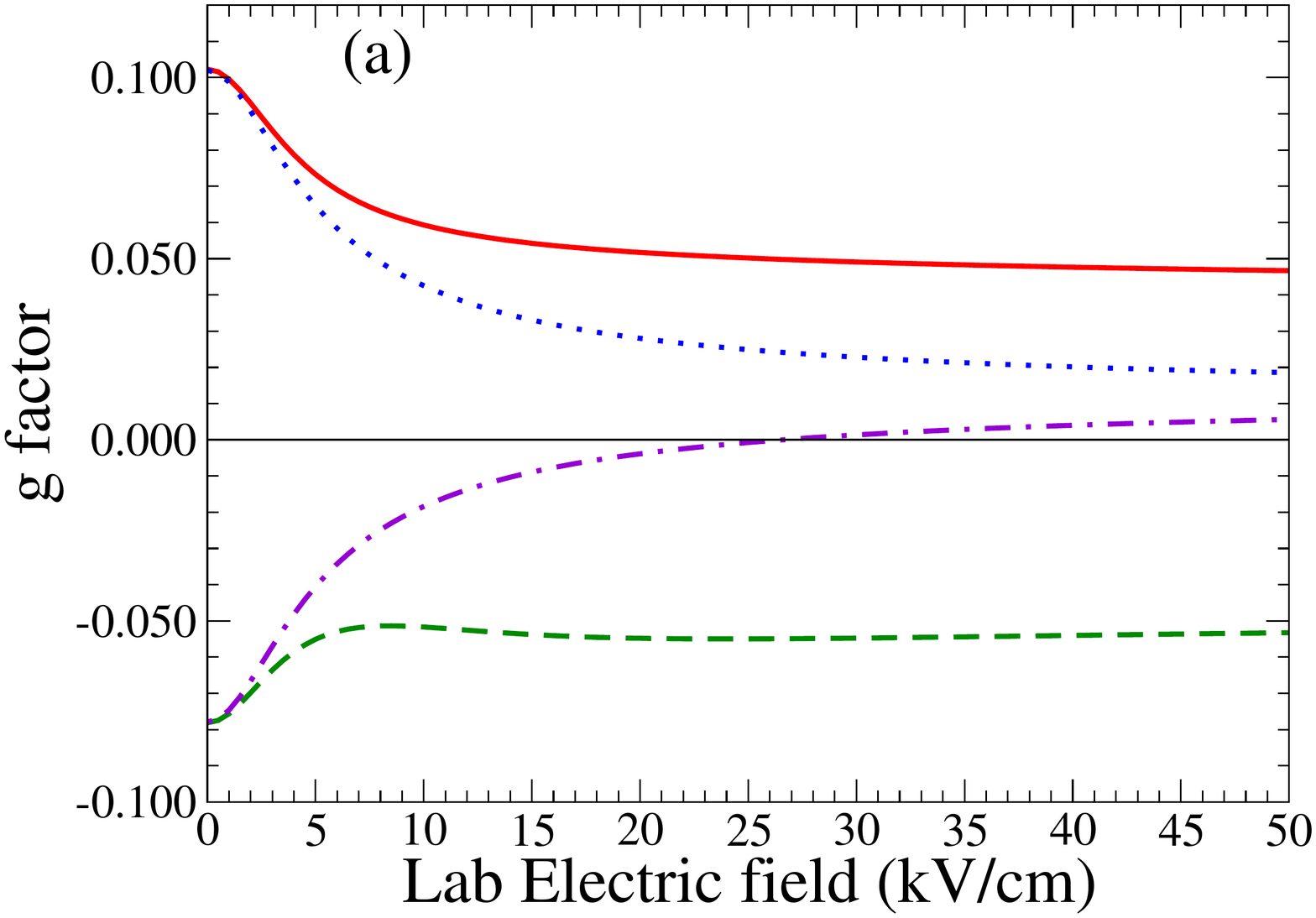}
\includegraphics[width=0.95\linewidth]{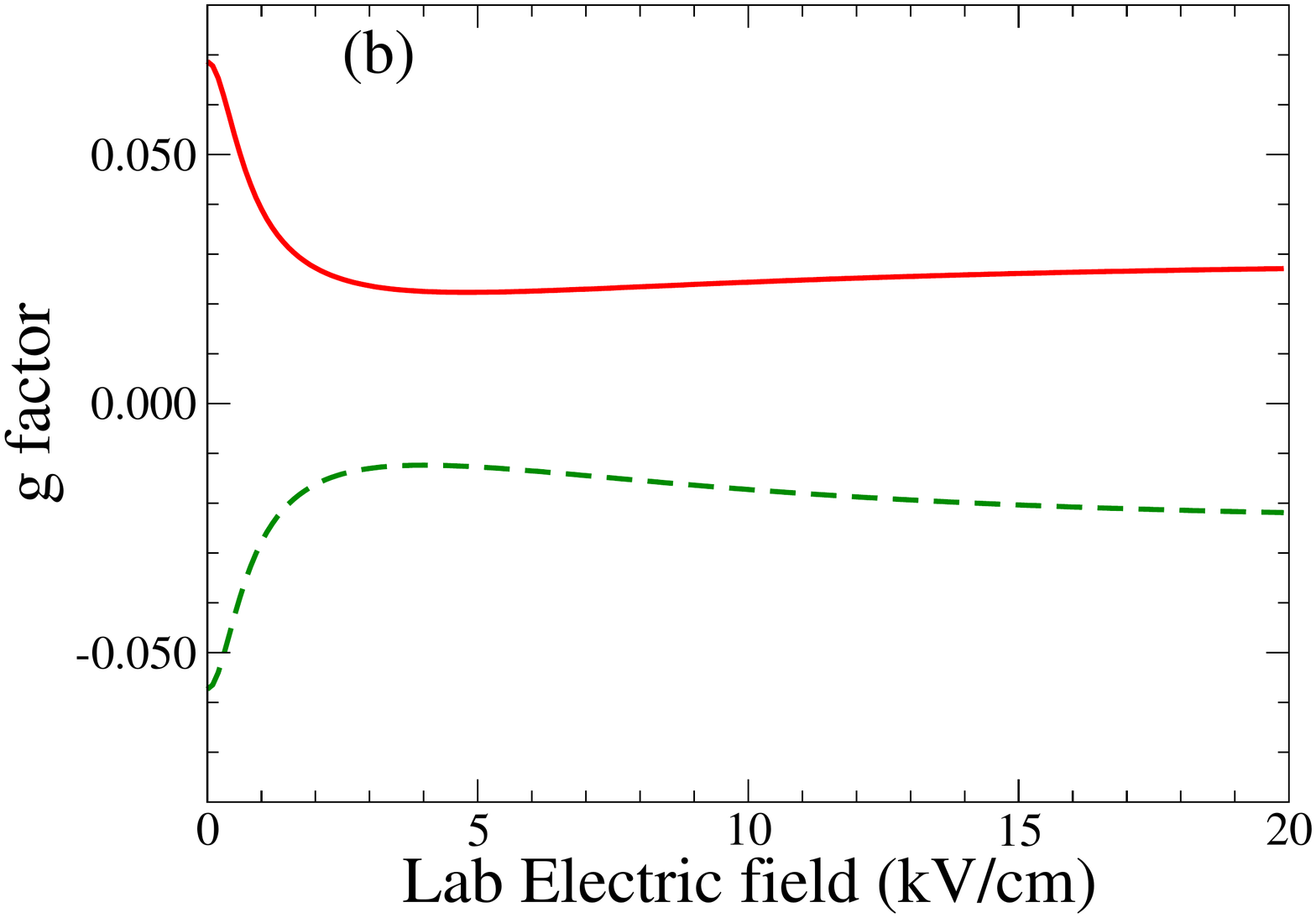}
\includegraphics[width=0.95\linewidth]{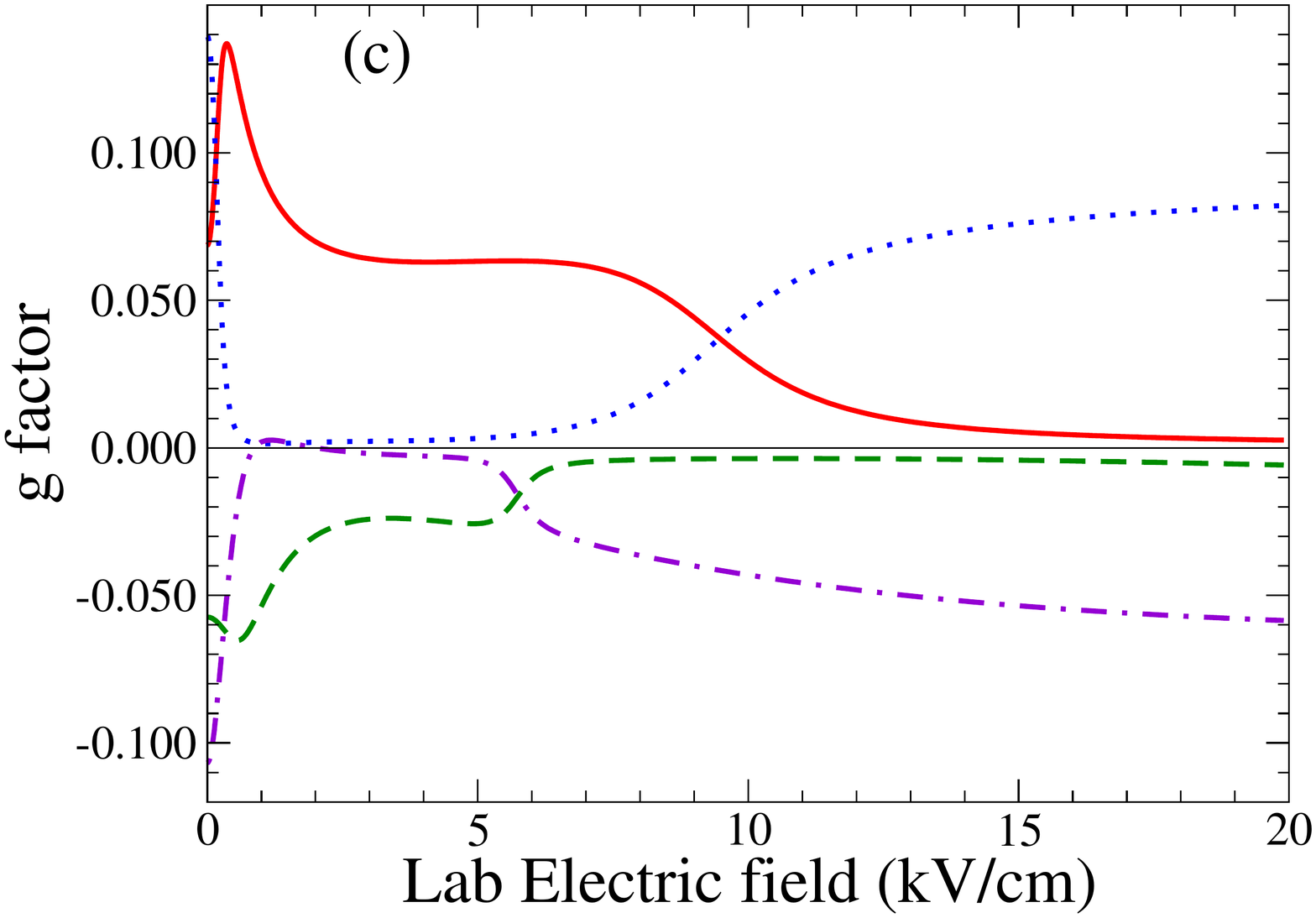}
\caption{\label{gfactor} 
Calculated $g$ factors.
(a) $^{208}$PbF. The solid (red) and dotted (blue) lines correspond to the $|M_F|$=1 lower-lying
$\Omega$-doublet states whereas the dashed (green) and the dashed-dotted (violet) lines correspond to the higher-lying $|M_F|$=1 states. The dotted (blue) and the dashed-dotted (violet) lines were calculated without
interaction with other rotational levels taken into account.
(b) $^{207}$PbF. The solid (red) line corresponds to the $|M_F|$=3/2 lower-lying
$\Omega$-doublet states whereas the dashed (green) line corresponds to the higher-lying $|M_F|$=3/2 states.
(c) $^{207}$PbF. The solid (red) line corresponds to the  lower-lying
$F$=3/2, $|M_F|$=1/2 $\Omega$-doublet states, the dashed (green) line corresponds to the higher-lying $F$=3/2, $|M_F|$=1/2 states, the dotted (blue) line corresponds to the lower-lying $F$=1/2, $|M_F|$=1/2 states,
 whereas the dashed-dotted (violet) line corresponds to the higher-lying $F$=1/2, $|M_F|$=1/2 states.}
\end{figure}
\endgroup

\section{Conclusions}
Experimental data and theoretical calculations for the $g$ factors of $^{207}$Pb$^{19}$F for the electric-field-free case are reported and found to be in a very good agreement with each other and with the body-fixed $g$ factors  $G_{\parallel}=0.081(5)$, $G_{\perp}=-0.27(1)$ obtained in Ref. \cite{Skripnikov:15d}. The calculated sensitivity to the electron electric dipole moment shows that an electric field of $1-2$ kV/cm is optimal for an experiment. The calculated electric-field-dependent $g$ factors provide the information needed to control systematic effects related to stray magnetic fields in future experiments such as those capitalizing on the coincidental near-degeneracy of levels of opposite parity in $^{207}$PbF.


\section*{Acknowledgments}
  The authors would like to thank A. L. Baum for his initial preparation of the experimental data and acknowledge Neil Shafer-Ray as a source of inspiration for this series of studies of PbF. Molecular calculations were supported by the Russian Science Foundation Grant No. 18-12-00227.  Work by T.J.S. was supported by the U.S. Department of Energy, Office of Science, Division of Chemical Sciences, Geosciences and Biosciences within the Office of Basic Energy Sciences, under Award No. DE-SC0018950. R.J.M. is grateful for research support provided by Pomona College
Sontag Fellowship and Hirsch Research Initiation Grant. P.M.R. is grateful for research support provided by SUNY-Oswego Office of Research and Sponsored Programs (ORSP). J.-U.G. acknowledges support from the Deutsche Forschungsgemeinschaft (DFG) Grants No. GR 1344/4-1, No. GR 1344/4-2, No. GR 1344/4-3, and No. GR 1344/5-1 and the Land Niedersachsen.
.

\bibliographystyle{apsrev}

\end{document}